\def \cm{~\rm{cm}}
\def \s{~\rm{s}}
\def \km{~\rm{km}}

\def \K{~\rm{K}}
\def \g{~\rm{g}}

\def \yr{~\rm{yr}}

\documentclass[referee]{aa}
\usepackage{graphicx}

\usepackage[german,english]{babel}
\begin{document}

%

\title{The absence of jets in cataclysmic variable stars\thanks{
Research supported by the Israel Science Foundation, and
the Centre National d'\'Etudes Spatiales} }

\author{Noam Soker\inst{1}
      \and Jean-Pierre Lasota\inst{2}}

\institute{Department of Physics, Technion-Israel Institute of
Technology,  32000 Haifa, Israel; soker@physics.technion.ac.il
\and
Institut d'Astrophysique de Paris, 98bis boulevard Arago, 75014
Paris, France; lasota@iap.fr}
\date{Received ---- / Accepted ----}

\titlerunning{Absence of jets in CVs}

\abstract{We show that the recently developed thermal model which
successfully describes how jets are launched by young stellar
objects, when applied to system containing disk-accreting white
dwarfs naturally explain the otherwise surprising absence of jets
in cataclysmic variable stars. Our main argument uses the crucial
element of the thermal model, namely that the accreted material is
strongly shocked due to large gradients of physical quantities in
the boundary layer, and then cools on a time scale longer than its
ejection time from the disk. In our scenario the magnetic fields
are weak, and serve only to recollimate the outflow at large
distances from the source, or to initiate the shock, but not as a
jet-driving agent. Using two criteria in that model, for the shock
formation and for the ejection of mass, we find the mass accretion
rate above which jets could be blown from accretion disks around
young stellar objects and white dwarfs. We find that these
accretion mass rates are $\dot M ({\rm YSO}) \ga 10^{-7} M_\odot
\yr ^{-1}$ and $\dot M ({\rm WD}) \ga 10^{-6} M_\odot \yr ^{-1}$
for young stellar objects and white dwarfs respectively.
Considering the uncertainties of the model, these limits could
overestimate the critical value by a factor of $\sim 10$. }
\maketitle


\section{Introduction} \label{sec:intro}

It is widely believed that most astrophysical jets, and all
massive jets (to distinguish from low density hot-plasma jets from
radio pulsars) are launched from accretion disks (Livio 1999,
2000a). This belief is supported by observations of jets in Young
Stellar Object (YSOs), Low-Mass and High-Mass X-ray Binaries
(LMXBs \& HMXBs) and Active Galactic Nuclei (AGNs) which all are
systems containing accretion disks (or at least accretion flows
with a considerable amount of angular momentum). The apparent
universality of the accretion disk--jet relation is spoiled by one
class of systems: Cataclysmic Variables (CVs). They are close
binary systems in which a white dwarf accretes matter lost by its
Roche-lobe filling low-mass companion (see Warner 1995 for a
review). For weak enough white-dwarf's magnetic fields CVs posses
accretion disks. But no jets have ever been observed from these
numerous and extremely well observed binaries. The one reported
occurrence (Shahbaz et al. 1997) failed to be confirmed (O'Brien
\& Cohen 1998) and the system itself is most probably not a CV but
a Super Soft X-ray Source (SSXS; Knigge, King \& Patterson 2000).
However, jets have been observed emanating from some other SSXSs
(see below) which shows that the absence of jets in CVs cannot be
attributed to some special properties of white dwarfs in binary
systems since white dwarfs are also present in SSXSs, and
symbiotic systems, some of which blow jets.

One can expect that the absence of jets in CVs could tell us
something about the still mysterious jets launching mechanism.
Commonly it is assumed that magnetic fields play crucial roles in
the formation of jets. Magnetic fields can appear in three
types of roles:

(1) In triggering the jet ejection events, e.g., by causing
instabilities in the disk. These types of ``magnetospheric" MHD
instabilities could exist in accretion disks even when the central
star has no magnetic field (e.g., Li \& Narayan 2004 and
references therein). MHD instabilities, turbulence, or other
disturbances may lead to strong shocks; the high post-shock
pressure may accelerate gas and form jets and/or winds, e.g., as
was shown for non-radiative accretion around a black hole by De
Villiers, Hawley, \& Krolik (2004).

(2) In accelerating the jets. There are many models and
countless of papers on this subject. Basically, most models are
based on the operation of large scale magnetic fields driving the
flow from the disk; either via the ``centrifugal wind'' mechanism,
first proposed by Blandford \& Payne (1982), or from a narrow
region in the magnetopause of the stellar field via an ``X-wind
mechanism'' introduced by Shu et al. (1988, 1991) and in a
somewhat different setting, by Ferreira \& Pelletier (1993, 1995).
See recent reviews by K\"onigl and Pudritz (2000), Shu et al.
(2000) and Ferreira (2002). It should, however, be pointed out
that the origin of the large scale magnetic fields and the manner
that open field lines of sufficiently strong magnitude
persist (in the centrifugal wind models), or the manner by which a
stellar field interacts with the disk, allowing inflow and at the
same time driving an outflow (in the X-wind models) are still open
key issues of the theory (e.g. Heyvaerts, Priest \& Bardou
1996). In addition, it seems that thermal pressure is needed in
some of these models (e.g., Ferreira \& Casse 2004).

(3) In collimating the jets (e.g., Heyvaerts \& Norman 1989). The
collimation issue is, however, still quite controversial and while
magnetic collimation is certainly plausible, its exact nature is
probably quite involved and still not fully understood (see the
recent works of Bogovalov \& Tsinganos 2001 and Li 2002).

Magnetic jet launching models fail to account for the absence
of jets in CVs despite of some interesting suggestions in Livio
(2000b). It is therefore justified to consider jet launching
mechanisms in which magnetic fields would be deprived of at least
one of the three roles. In the present article we show that the
absence of jets in CVs is naturally explained by the model of
thermal pressure acceleration proposed by Torbett (1984)
and Torbett \& Gilden (1992), and which was developed
and extended recently by Soker \& Regev (2003; hereafter SR03)
to explain collimated outflows in YSO.

SR03 interest in thermal pressure acceleration of jets was
motivated by new results from recent X-ray observations of YSO.
These show that there is essentially no difference between the
properties of X-ray emission from YSO with and without outflows
(Getman et al.\ 2002), imposing quite severe constraints on models
based on magnetic launching of jets. In the thermally-driven
jet model, the magnetic fields are weak, and serve only to
re-collimate the outflow at large distances from the source, (role
(3) above), and possibly trigger disturbances in the boundary
layer (BL), (role (1) above). The BL is the inner layer, where the
disk adjust itself to the conditions of the accreting star. A
crucial ingredient of the model is that the accreted material is
strongly shocked, and that it cools down on a time scale longer
than its ejection time from the inner disk. SR03 find that the
thermal acceleration mechanism works only when the accretion rate
in YSO accretion disk is large enough and the $\alpha$ parameter
of the disk small enough - otherwise the cooling time is too short
and significant ejection does not take place. SR03 term these
strong shocks `spatiotemporally localized (but not too small!)
accretion shocks', or SPLASHes.

In the present paper we extend the analysis of SR03 to white
dwarfs (WDs) accretors. We compare the derived conditions for
thermally launching jets from accretions disks around YSO and
around WDs. YSO refers also to main sequence stars accreting from
mass-losing companions stars. We show that the thermally-driven
jet model can be extended to jets blown by disks around WDs and
explain the absence of jets in CVs.

\section{The Opacity Mechanism} \label{sec:opac}

There are two basic radiative cooling time-scales, which lead to
two conditions for launching of strong outflows by thermal
pressure. The first is that the photon-diffusion cooling time of
the entire BL region, $\tau_{\rm diff}$, be longer than the
ejection time of the jet ${\tau_{\rm ej}}$. This applies to the
jet-acceleration phase, after the disk-material has passed through
a strong shock. In other words, this is the condition for the
SPLASH to be able to eject shocked material. The second time scale
refers to the buildup phase of a strong shock, i.e. determines the
condition for the SPLASH formation. In the following sections we
will consider the constrains on these two cooling time-scales,

The photon diffusion time-scales depend on the opacity, which
itself strongly depends on the density and temperature (e.g.,
Rogers \& Iglesias 1992; Seaton et al. 1994). In Figure (1) we
plot the mean Rosseland opacity coefficient as a function of
the temperature for four values of the density. The opacity is
taken from Seaton et al.\ (1994), for a composition of $X=0.7$ and
$Z=0.03$. We take a little higher than solar metallicity to
account for enrichment in presently formed YSOs.

The temperature range characterizes accretion disks and BLs around
YSOs is $\sim 10^4 - 10^6 \K$, while around a WD accretor this
range is $\sim 10^6 - 10^8 \K$. From Figure 1 it appears
immediately that the opacity behaves markedly different in these
two ranges. We will explore the significance of this
difference in the following sections.
\begin{figure}
{\includegraphics [width=13.5cm] {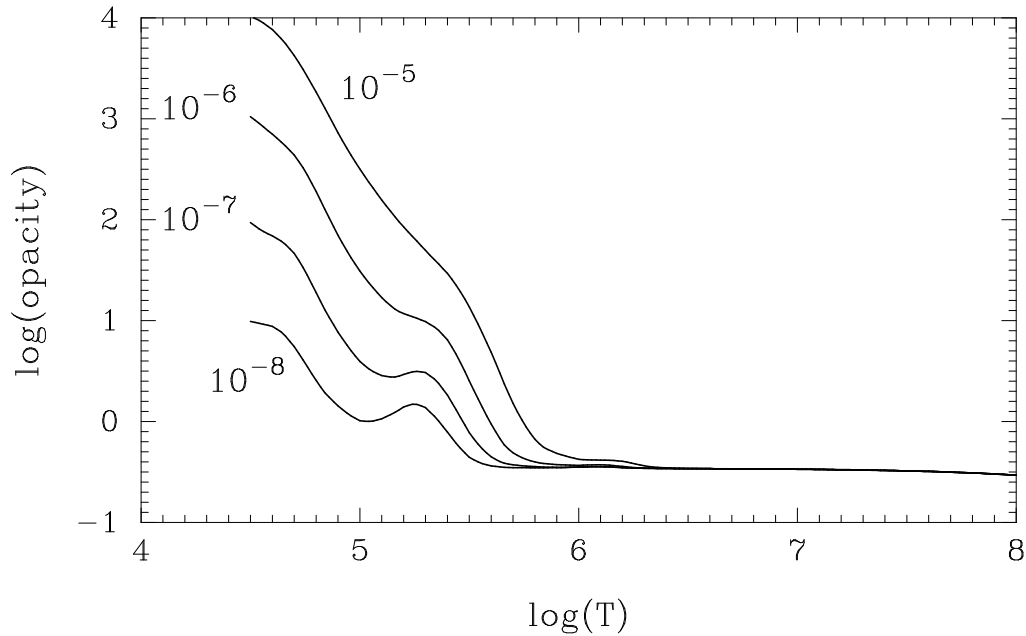}} \vskip 0.2 cm
\caption{Opacity as a function of temperature for four values of
the density as indicated, and for a composition of $X=0.7$ and
$Z=0.03$. Quantities are given in cgs units.}
\end{figure}
The opacity coefficient in the relevant range for YSO
accretors can be adequately fitted by
\begin{equation}
\log \kappa
=1+0.7\log\left(\frac{\rho}{10^{-7}}\right)
-2\log\left(\frac{T}{10^{5}} \right) -
0.6\log\left(\frac{\rho}{10^{-7}}\right)
\log\left(\frac{T}{10^{5}} \right),
\label{kappa1}
\end{equation}
where $\kappa$ is the opacity coefficient and $\rho$ the
density in the disk; all quantities being expressed in cgs units.
For later calculations it will be convenient to write the opacity
as
\begin{equation}
\kappa=10 \left(\frac{\rho}{10^{-7}}\right)^{0.7}
\left(\frac{T}{10^{5}} \right)^{-2}
\left(\frac{\rho}{10^{-7}}\right)^ {-0.6 \log(T/10^{5})} .
\label{kappa2}
\end{equation}
The last two approximations hold for
$-8 \la \log \rho \la -5$ and
$4.5 \la \log T \la 5.5$.

\section{The acceleration phase} \label{sec:accel}

The radiative diffusion time from the entire BL region depends on
the conditions in the disk BL and can be determined from the
photon mean free path $\lambda$, and the size of the region from
which the photons have to escape. It is assumed that the gas has
been strongly shocked. Using formulae from Torbett (1984),
Torbett \& Gilden (1992) and  SR03, one can write the
diffusion time as
\begin{equation}
\tau_{\rm diff} = \epsilon^2\, R^2\, \frac{\rho \kappa}{c},
\label{taudiff1}
\end{equation}
where $\epsilon = H/R$, $H$ is the vertical disk's scale height,
$R$ is the distance from the center of the disk,
and $c$ is the speed of light.
We will use the expression for
the density in an $\alpha-$disk model in which the radial velocity
of the slowly accreting mass is given by (Pringle 1981) $v_r
\simeq \alpha \epsilon^2 v_{\rm K}$, where $\alpha$ is the
disk-viscosity parameter, and $v_{\rm K}$ the Keplerian tangential
velocity. From the mass conservation equation $\dot M = 2 \pi R 2H
\rho v_r$, one finds
\begin{equation}
\rho=
\frac{\dot M}{4\pi \alpha \epsilon^3 R^2 v_K}
\label{dens1}
\end{equation}
Taking into account that behind a strong shock $\rho$
increases by a factor four, one gets for the diffusion time
\begin{equation}
\tau_{\rm diff}=\frac{\dot M \kappa}{\pi c \alpha  \epsilon v_K}
\label{taudiff2}
\end{equation}
This has to be compared with the dynamical ejection time
$\tau_{\rm ej} = H/v_{\rm esc}= \epsilon R /v_{\rm esc}$.
Taking $v_K \simeq 0.7 v_{\rm esc}$ one has
\begin{equation}
\frac {\tau_{\rm diff}}{\tau_{\rm ej}} \simeq \nonumber
\frac{\dot M \kappa}{\pi c \alpha  \epsilon^2 R} \\
\simeq 1.3
\left( \frac{\epsilon}{0.1} \right)^{-2}
\left(\frac{\alpha}{0.1} \right)^{-1}
\left(\frac{\dot M}{10^{-7} M_\odot \yr^{-1}}\right)
\left(\frac{\kappa}{\rm cm^2 \g^{-1}} \right)
\left(\frac{R} {R_{\sun}} \right)^{-1}.
\label{taudiff3}
\end{equation}

For a SPLASH to be able to eject a collimated outflow
the ejection time $\tau_{\rm ej}$ must be shorter than the
radiation diffusion time $\tau_{\rm diff}$ (SR03). For the strong
shocks the temperature is $T \ga 10^6$ (see below), and $\kappa
\simeq 0.4$, for both YSOs and WDs. The condition $\tau_{\rm diff}
\ga \tau_{\rm ej}$  is met for YSOs, $R \simeq 1 R_\odot$,
accreting at a rate of
\begin{equation}
\dot M_{s} \ga 2 \times 10^{-7} \left( \frac{\epsilon}{0.1}
\right)^{2} \left( \frac{\alpha}{0.1} \right) M_\odot \yr^{-1}
\qquad {\rm for} \quad {\rm YSOs},
\label{accyso1}
\end{equation}
whereas for WDs ($R\simeq 0.01 R_\odot$) it requires
\begin{equation}
\dot M_{s} \ga
2 \times 10^{-9}
\left( \frac{\epsilon}{0.1} \right)^{2}
\left( \frac{\alpha}{0.1} \right)
M_\odot \yr^{-1} \qquad {\rm for} \quad {\rm WDs},
\label{accwd1}
\end{equation}
which means that for both types of objects if a SPLASH forms it
would be able to eject collimated flows. We turn now to the second
cooling time-scale which determines when SPLASH can form.

\section{The buildup of a strong shock} \label{sec:build}

The SPLASH model assumes that hundreds of small blobs are
formed in the sheared BL (section 2 of SR03). The blobs
occasionally collide with each other, and create shocks which
cause the shocked regions to expand in all directions. If the
shocked regions continue to expand out into the path of yet more
circulating blobs, stronger shocks may be created, as was proposed
by Pringle \& Savonije (1979) to explain the emission of X-rays
out of disk BLs in dwarf novae. For the shocked blobs to expand,
the radiative cooling time of {\em individual blobs}, $t_{\rm
cool} \simeq \ell^2 \kappa \rho_b/c$, must be longer than the
adiabatic expansion time of individual blobs $t_{\rm ad} =
\ell/c_s$, where $\ell$ is the size of an expanding blob, and
$\rho_b$ the post-shock blob's density. SR03 demand also that the
blobs be small, because the dissipation time of disk material to
form the SPLASH must be shorter than the jet ejection time (eq. 24
of SR03). This gives for the blobs size $\ell \sim \epsilon^2 R$
(eq. 27 of SR03). The condition of long cooling time for
individual blobs becomes (see eqs. 29 and 30 of SR03)
\begin{equation}
\ell \kappa \rho_b \ga \frac{c}{c_s},
\label{blob1}
\end{equation}
where $c_s$ is the isothermal sound speed, that is, the
optical depth of the blob must be larger than the ratio of light
to sound speed. Scaling the variables in the last condition with
$\ell=\epsilon^2 R$, gives,
\begin{equation}
1 \la \eta \equiv \epsilon^2 \kappa R \rho_b \frac{c_s}{c},
\label{blob2}
\end{equation}
which defines the function $\eta$. Substituting the density
from equation ({\ref{dens1}}) in equation ({\ref{blob2}}) gives
\begin{equation}
1 \la \eta = \frac{\dot M \kappa}{4 \pi c \alpha R \epsilon}
\left(\frac{c_s} {v_K}\right)
\simeq
\frac{\dot M \kappa}{4 \pi c \alpha R},
\label{blob3}
\end{equation}
since $c_s/v_K \simeq \epsilon$. Therefore the condition
$\eta \ga 1$ requires
\begin{equation}
\dot M \ga 4.2 \times 10^{-5}
\kappa^{-1}
\left( \frac{\alpha}{0.1} \right)
\left(\frac{R} {R_{\sun}} \right)
M_\odot \yr^{-1} .
\label{acc01}
\end{equation}

The weak shock temperature $T_b$ will be intermediate between the
strong shock temperature
\begin{equation}
T_s = \frac{3}{16}\frac{\mu m_H}{k}
v_K^2
\simeq
1.4 \times 10^{-9}v_K^2 \simeq 2.6 \times 10^6
\left(\frac{M} {\rm M_{\odot}} \right)
\left(\frac{R} {R_{\sun}}\right)^{-1}\rm K
\label{shockt}
\end{equation}
and the disk temperature, $T_b \simeq 0.05 T_s$, say.

The post-shock temperature of the blobs can be a little higher
than $T_b=5 \times 10^4 \K$ for YSOs and higher than $T_b = 5
\times 10^6 \K$ for WD accretors. The cooling time from higher
temperatures is shorter. However, what is important for the model
is that the shocked blobs don't cool back to the original
temperature of $T \sim 10^4 \K$ for YSO and $T \sim 10^6$ for WD
accretors, so that an expansion process of the BL region starts.

For an YSO accretor we take the opacity as given in equation
({\ref{kappa2}}). Substituting $T_b=5 \times 10^4 \K$ gives
$\kappa= 40(\rho/10^{-7})^{0.88}$. We take the density from
equation ({\ref{dens1}}) scaled to YSOs, and then substitute
the opacity in equation ({\ref{acc01}}). The condition for the
formation of a strong shock from weakly-shocked small blobs
({\ref{acc01}}) becomes a condition on the accretion rate
\begin{equation}
\dot M_b \ga
7 \times 10^{-7}  
\left( \frac{\epsilon}{0.1} \right)^{1.4}
\left( \frac{\alpha}{0.1} \right)
M_\odot \yr^{-1} \qquad {\rm for} \quad {\rm YSOs}.
\label{accyso2}
\end{equation}

For WD accretors we take $T_b > 5 \times 10^6$, $R=0.01
R_\odot$. From Figure (1) it is clear that the opacity is
$\kappa=0.4 \cm^2 \g^{-1}$. The condition of equation
({\ref{blob3}}) then reads
\begin{equation}
\dot M_b \ga
10^{-6}
\left( \frac{\alpha}{0.1} \right)
M_\odot \yr^{-1} \qquad {\rm for} \quad {\rm WDs},
\label{accwd2}
\end{equation}
which is never satisfied in CVs (Warner 1995).

Can a shock be built away from the accreting body, and launch jets
there? First from equations ({\ref{taudiff3}}) and ({\ref{acc01}})
we note that the conditions for accelerating the gas and building
a SPLASH are more difficult to meet away from the accreting
object. From both equations the accretion rate should increase as
$R$. Second, and more important, to build a shock a strong
perturbation should occur in the disk. This can be triggered by
large gradient in one or more of the physical variables. This is
naturally the case in the boundary layer (and perhaps close
to the last stable orbit around black holes; this is outside the
scope of the present paper). In any case, we cannot rule out the
possibility that jets will be launched somewhat away from the
boundary layer, up to $\sim 3$ times the boundary layer radius.

\section{Discussion and summary} \label{sec:discuss}

In the previous two sections we found the constraints on the  mass
accretion rate derived using the two conditions for slow radiative
cooling derived by SR03. The first condition is that the strongly
shocked gas in the BL will cool slowly, such that the thermal
pressure will have enough time to accelerate the jet's material.
The constraints on the accretion rate, $\dot M_s$, are given by
equations ({\ref{accyso1}}) and ({\ref{accwd1}}), for YSO and WD
accretors, respectively. The second condition is that weakly
shocked blobs in the BL will expand, and disturb the BL in such a
way that a strong shock will develop. The constraints on the
accretion rate, $\dot M_b$, are given by equations
({\ref{accyso2}}) and ({\ref{accwd2}}), for YSO and WD accretors,
respectively. We should stress the following in regard to these
constraints. (1) The constraints are accurate to an order of
magnitude. This is for several reasons, e.g.,  the values of
$\alpha$ and $\epsilon$ are unknown; the demand on the ratio of
the cooling time to acceleration time is given to an order of
magnitude; the behavior of the gas, e.g., its opacity and
pressure, should be treated more accurately with a full 3D
numerical code. (2) The first constraint on $\dot M_s$ is
generic to the proposed thermally-accelerated jet model. (3) The
second one, on the formation of the disturbances that lead to the
formation of strong shocks in the BL, might be less important.
This is because other types of disturbances can cause strong
shocks to develop in the BL, e.g., MHD instabilities, magnetic
eruptions from YSOs, and local thermonuclear events on a WDs.

With these, we note the following. For YSOs (and other main
sequence stellar accretors)  the two requirements on the two
cooling time-scales basically gives the same constraint on the
mass accretion rate $\dot M({\rm YSO}) \ga 10^{-7} M_\odot
\yr^{-1}$ (eqs. {\ref{accyso1}} and {\ref{accyso2}}). This fits
observations, which show jets from YSO accreting at such rates
(e.g., Cabrit et al.\ 1990).

For WDs the two constraints give $\dot M_s \ga 10^{-9} M_\odot
\yr^{-1}$ and $\dot M_b \ga 10^{-6} M_\odot \yr^{-1}$ (eqs.
{\ref{accwd1}} and {\ref{accwd2}}). The second one is very
stringent. The highest accretion rates deduced from
observations of nova-like systems or dwarf-novae at
maximum are two orders of magnitude lower (see e.g. Warner 1995).
Taking into account the order of magnitude uncertainty one could
maybe get $\dot M({\rm WD}) \ga 1-3 \times 10^{-7}$. This is
compatible with observations of SSXSs and symbiotic systems. SSXSs
are thought to be white dwarfs accreting at rates of $3 \times
10^{-8}-10^{-6} M_\odot \yr^{-1}$ from a companion, and sustaining
nuclear burning on their surface (e.g., van den Heuvel et al.\
1992; Greiner 1996). To maintain a steady nuclear burning the mass
accretion rate should be $3 \times 10^{-8}-10^{-6} M_\odot
\yr^{-1}$, where the upper range is for massive WDs (Nomoto 1982).
Fast, $\sim 1000-5000 \km \s^{-1}$, collimated outflows have been
observed in some SSXSs, RX J0513.9-6951 (Crampton et al.\ 1996;
Southwell et al.\ 1996), RX J0019.8+2156 (Becker et al. 1998;
Quaintrell \& Fender 1998; Tomov et al.\ 1998), and RX
J0925.7-4758 (Motch 1998). These systems teach us that WDs
accreting mass at rates much higher  than those in cataclysmic
variables can blow jets. In RX J0925.7-4758 the high luminosity
(Motch 1998) implies an accretion rate of $\sim 10^{-7} M_\odot
\yr^{-1}$, and the WD radius is $\sim 0.005 R_\odot$. With
$\epsilon = 0.05$ the constraint (eq. {\ref{accwd2}}) is $\dot M_b
\ga 3 \times 10^{-7} M_\odot \yr^{-1}$. The accretion rate onto RX
J0513.9-6951 is even higher (Southwell et al.\ 1996). Some
symbiotic systems are also known to blow jets (Sokoloski 2004;
Brocksopp et al.\ 2004 and references therein). In these symbiotic
systems, WD companions accrete from the wind of red giant branch
stars or asymptotic giant branch stars, at relatively high rates.
In some of the symbiotic systems which blow jets the WD sustains a
quasi-steady nuclear burning, similar to SSXSs; in others, there
is no nuclear burning (Brocksopp et al.\ 2004). Still, the mass
accretion rate is expected to be high in the later group as well.
It is possible that in the later systems the WD are more massive;
more massive WD stars require higher mass accretion rates to
sustain nuclear burning (Nomoto 1982).

Although no jets have ever been observed in CVs some of them emit
winds. P Cygni profiles in resonant UV lines are observed in some
very luminous CVs such as the nova-like stars and dwarf novae at
outburst maximum. These winds are too cold to be ejected by a
thermal mechanisms. They are most probably radiation driven
(see e.g., Murray, 2002; Proga 2002).

Finally, it is interesting to investigate what kind of condition
one obtains when considering ultra-compact objects such as neutron
stars and black holes. In such a case it is more convenient to
scale Eq. ({\ref{acc01}}) with the Schwarzschild radius $R_{\rm
G}=2GM/c^2$ and the Eddington accretion rate $\dot M_{\rm Edd}=
L_{\rm Edd}/0.1c^2= 2.3\times 10^{-8} M_\odot \yr^{-1}$. One
obtains then
\begin{equation}
\dot m \equiv \frac{\dot M}{\dot M_{\rm Edd}}
\ga 0.02
\left( \frac{\alpha}{0.1} \right)
\left( \frac{0.4 \rm \ g \ cm^{-2}}
{\kappa}\right)
\frac{R}{R_{\rm G}},
\label{comp}
\end{equation}
which is very close to the value at which low-mass X-ray
transients enter (from below) the so called hard/low states
associated with the appearance of steady jets (e.g., Fender 2001).
Of course there are no boundary layers around accreting black
holes so that our model cannot be directly applied to these
objects. We will discuss this problem in a future paper.

\acknowledgements We thank Marek Sikora for his precious remarks
on the manuscript. JPL is grateful to Daniel Proga for very useful
discussions.

{}

\end{document}